\documentclass[
aps,%
12pt,%
final,%
notitlepage,%
oneside,%
onecolumn,%
nobibnotes,%
nofootinbib,% In the current version of REVTeX, when this option is on,
superscriptaddress,%
noshowpacs]%
{revtex4}
\RequirePackage{graphicx}

\def\refitem#1{\relax}

\begin{document}
\title{Debye mass and heavy quark potential in a PNJL quark plasma}

\author{\firstname{J.} \surname{Jankowski}}
\email{jakubj@ift.uni.wroc.pl}
\affiliation{Institute for Theoretical Physics, University of Wroclaw, 
50-204 Wroclaw, Poland}

\author{\firstname{D.} \surname{Blaschke}}
\email{blaschke@ift.uni.wroc.pl}
\affiliation{Institute for Theoretical Physics, University of Wroclaw, 
50-204 Wroclaw, Poland}
\affiliation{Bogoliubov  Laboratory of Theoretical Physics, JINR Dubna, 
141980 Dubna, Russia}

\begin{abstract}
We calculate the Debye mass for the screening of the heavy quark potential 
in a plasma of massless quarks coupled to the temporal gluon background 
governed by the Polyakov loop potential within the PNJL model in RPA 
approximation.
We give a physical motivation for a recent phenomenological fit of lattice 
data by applying the calculated Debye mass with its suppression in the 
confined phase due to the Polyakov-loop to a description of the temperature
dependence of the singlet free energy for QCD with a heavy quark pair at 
infinite separation.
We compare the result to lattice data.
\end{abstract}

\maketitle

\section{Introduction}
The proposal of Matsui and Satz \cite{Matsui:1986dk} that
color screening of heavy quarks in a deconfined medium 
should lead to a dissociation of the $J/\psi$ 
bound state and whence could be a 
clear signal of quark-gluon plasma (QGP)
formation in heavy-ion collision (HIC) experiments
arose to a wide field of research both in 
experimental as well as theoretical directions \cite{Rapp:2008tf}.
Following the original idea the simplest  picture is to 
consider a static quark-antiquark ($Q\bar{Q}$) probe
in a color plasma where all medium effects are included in
two body interaction potential $V_T(r)$. 
Then solving the $Q\bar{Q}$ $T$-matrix \cite{Riek:2010fk}     
for such a system (or equivalently Schr\"{o}dinger equation 
\cite{Ebeling:1986}) one can obtain a temperature modification 
of the hadron spectra. 
For obvious reasons this description is unrealistic 
(e.g., in HIC experiments the medium always has collective flow) 
but nevertheless it can provide physically reliable insights.
But what potential should we use? 
One possibility of approaching this problem is to use nonperturbative
field theoretical methods to motivate an ansatz
for the heavy quark potential and then fit its parameters
to the lattice data. 
For example in ref. \cite{Megias:2007pq,Riek:2010fk}
it has been shown that including effects of a dimension-two 
gluon condensate one gets quarkonium spectral
functions in good agreement with lattice data. 

In this note we go similar way by investigating Debye screening
effects in modifying the vacuum Cornell potential (one-gluon exchange 
plus linear term) when going to a finite temperature medium. 
The screening medium is composed of massless quarks coupled to
the homogeneous, temporal gluon background which is 
a very crude approximation of confinement. 
This essentially constitutes a version of the PNJL model 
\cite{Ratti:2005jh,Hansen:2006ee} which was found to reproduce lattice 
thermodynamics although it is theoretically still rather unsatisfactory.
In this contribution we discuss the similarities of the temperature 
dependence of the Debye mass within such a model and that obtained from
a fit of lattice QCD heavy quark free energy with the effective interaction of 
Ref.~\cite{Riek:2010fk}. 
This observation holds promise for a better microscopic understanding 
of the physical effects determining the behaviour of the heavy quark-antiquark 
potential at finite temperature.

%%%%%%%%%%%%%%%%%%%%%%%%%%%%%%

\section{Screening in the gluon background}

\begin{figure}
\includegraphics[width=0.5\textwidth]{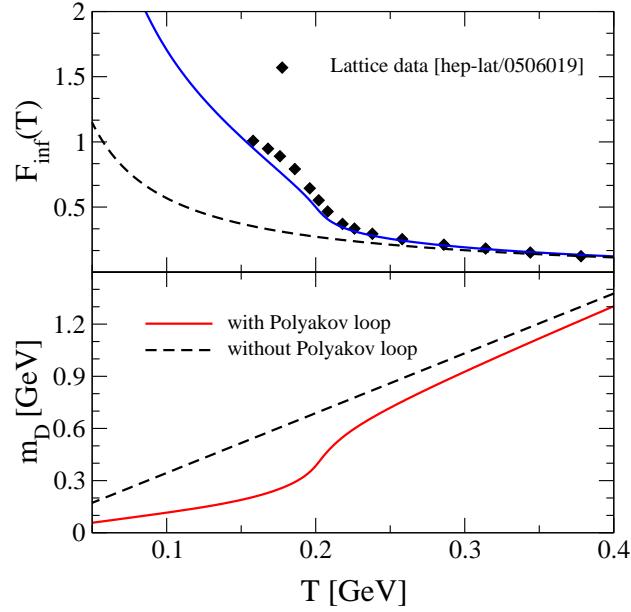}
\caption{Lower panel: Temperature dependence of the Debye mass 
with (solid line) and without (dashed line) inclusion of Polyakov loop. 
Calculated for $\alpha_s=0.471$.
Upper panel: Temperature dependence of $F_1(r=\infty,T)$  with (solid line) 
and without (dashed line) inclusion of Polyakov loop, compared to lattice data 
from Ref.~\cite{Kaczmarek:2005gi}.
}
\label{Fig.1}
\end{figure}

In Ref.~\cite{Jankowski:2009kr} a model was considered 
where the vacuum potential was screened by quark matter polarization 
decribed by quark-antiquark loop integrals where the internal lines 
were coupled to a temporal gluon background field.
Specifically, when the static interaction potential is given as
$V(q)$, $ q^{2} = |{\bf{q}}|^{2} $, 
the statically screened potential is given by a resummation of 
one-particle irreducible diagrams ("bubble" resummation = RPA)
\begin{equation}
V_{\rm sc}(q) = {V(q)}/[{1 - \Pi_{00}(0; {\bf q})/q^{2}}]~,
\label{Vsc}
\end{equation}
where the longitudinal polarization function 
%$ F(0; {\bf q}) = - \Pi_{00}(0; {\bf q}) $ 
in the finite $ T $ case can be calculated within 
thermal field theory as
\begin{equation}
\Pi_{00}(i\omega_{l};{\bf q} ) 
= g^{2} T\sum_{n=-\infty}^{\infty} \int\frac{d^{3}p}{(2\pi)^{3}} 
{\textrm{ Tr}} [\gamma^{0}S_{\Phi}(i\omega_{n}; {\bf p})
\gamma^{0}S_{\Phi}(i\omega_{n}-i\omega_{l}; {\bf p} - {\bf q})]~.
\end{equation}
Here $\omega_{l}=2\pi lT$ are the bosonic and $\omega_{n}=(2n+1)\pi T$
are the fermionic Matsubara frequencies of the imaginary-time formalism.
The symbol ${\textrm{Tr}}$ stands for traces in color, flavor and Dirac 
spaces.
$S_{\Phi}$ is the propagator of a massless fermion coupled to the homogeneous 
static gluon background field $\varphi_3$. Its inverse is given by 
\cite{Ratti:2005jh,Hansen:2006ee}
\begin{equation}
S^{-1}_{\Phi}( {\bf p}; \omega_{n} ) = 
{\bf \gamma\cdot p} + \gamma_{0}i\omega_{n} -\gamma_0\lambda_{3}\varphi_3~,
\label{coupling}
\end{equation}
where $\varphi_3$ is related to the Polyakov loop variable defined by
\cite{Ratti:2005jh}
$$ \Phi(T) = \frac{1}{3}\rm Tr_c (e^{i\beta\lambda_{3}\varphi_{3}}) 
= \frac{1}{3}(1 + 2\cos(\beta\varphi_3) )~. $$ 
The physics of $\Phi(T)$ is governed by the temperature-dependent Polyakov 
loop potential ${\cal{U}}(\Phi)$, which is fitted to describe the lattice data 
for the pressure of the pure glue system  \cite{Ratti:2005jh}. 
After performing the color-, flavor- and Dirac traces and making the fermionic 
Matsubara summation, we obtain in the static, long wavelength limit 
\begin{eqnarray}
\Pi_{00}( {\bf q} )  
= \frac{2N_cN_f g^2}{\pi^2} \int_{0}^{\infty}dp\, 
p^{2}\frac{\partial f_\Phi}{\partial p} 
%\nonumber
%= - \frac{ 4N_{\rm dof}g^2}{\pi^2} \int_{0}^{\infty}dp\,p f_{\Phi}(p) 
= -2 g^{2}T^{2}I(\Phi) = - m_{D}^2(T) ~,
\label{debyemass}
\end{eqnarray}
where $m_{D}(T)$ is the Debye mass, the number of degrees of freedom 
is $N_c=3$, $N_f=2$ and $f_\Phi(p)$ is the quark distribution 
function \cite{Hansen:2006ee}. 
In comparison to the free fermion case \cite{LeBellac,Beraudo:2007ky} the 
coupling to the Polyakov loop variable $\Phi(T)$ gives rise to a modification 
of the Debye mass, given by the integral
\begin{equation}
I(\Phi) = \frac{12}{\pi^2}\int_{0}^{\infty}\,
dx\,x\frac{\Phi(1+2e^{- x})e^{- x}+e^{-3 x}}
{1 + 3\Phi(1 + e^{- x})e^{- x}+e^{-3 x}}.
\end{equation}
The temperature dependence of $\Phi(T)$ is taken from 
%the nonlocal PNJL model of 
Ref.~\cite{Blaschke:2007np}.
In the limit of deconfinement ($\Phi = 1$), the case of a massless
quark gas is obtained ($I(1)=1$), while for confinement ($\Phi = 0$) one finds
that $I(0)=1/9$. 
The temperature dependence of the resulting Debye mass is shown in 
Fig.~\ref{Fig.1}  and as expected from the very beginning is much lower,
comparing to the free, massless case, in the confined and
transition region (with the critical temperature $T_c\approx200~\textrm{MeV}$).
For temperatures $T>>T_c$ the free gas behavior is reproduced.
Adopting the field theoretical approach of Ref.~\cite{Megias:2007pq,Riek:2010fk}
that  the color singlet free energy at finite temperature 
is driven by the screened, nonperturbative gluon propagator
and that Debye masses in Coulomb and stringy
sectors are different, we get ($\alpha_s=g^2/4\pi$)
\begin{equation}
\label{Vs}
F_{\rm \bar{Q}Q}(r) = -\frac{4}{3}\alpha_s\left(
\frac{e^{-m_Dr}}{r} - \frac{m_G^2}{2\widetilde{m}_D} 
+ \frac{m_G^2}{2\widetilde{m}_D}e^{-\widetilde{m}_Dr} + m_D 
\right)~,
\label{FE}
\end{equation}
where the constant ($r$- independent) term is a homogeneous mean
field contribution and takes into account 
the one-particle self-energy effects
\cite{Rapp:2008tf,Ebeling:1986}.
We take $m_G^2=0.631~\textrm{GeV}^2$ which gives 
string tension $\sigma=2m_G^2\alpha_s/3=0.198~\textrm{GeV}^2$.
We notice that our calculated Debye mass resembles 
qualitatively the behavior resulting from a recent fit for the stringy term 
in Ref.~\cite{Riek:2010fk} with the difference that our result is smaller than 
that of the fit by abut a factor two. 
Therefore we proceed by plugging it in for $\widetilde{m_D}$ in equation 
(\ref{FE}), and if we define
\begin{equation}
F_\infty(T)=F_{\bar{Q}Q}(r\rightarrow\infty)
=\frac{4}{3}\alpha_s\left(\frac{m_G^2}{2\widetilde{m}_D}-m_D\right)~.
\label{eq:Finfty}
\end{equation}
Then we see that $F_\infty(T\rightarrow\infty)\approx-T$, 
and that it diverges for small temperatures - faster than in
the free fermion case due to gluon suppression of $\widetilde{m}_D$ - 
which is expected in quenched approximation but not in
QCD with dynamical quarks which would make $F_\infty$
finite due to so called "string breaking mechanism".
Fitting lattice data for $F_\infty(T)$ \cite{Kaczmarek:2005gi}
for high temperatures results in a coulombic Debye mass $m_D=0.12~T$
which is much lower than the perturbative one which is about $3.143~T$
for two flavors and $\alpha_s=0.471$.
We plot the resulting free energies in Fig.~\ref{Fig.1}. 
What concerns $F_\infty(T)$ we see that our results meet quite well
the lattice data in a quite a wide temperature range
starting from the transition region up to high temperatures;
the generic shape  is reproduced
in the transition region, while we see a discrepancy for low temperatures.
Our result is smaller than the lattice (especially for low $T$)
which is a remnant of our lower Debye mass than the one 
of the fit by Riek and Rapp \cite{Rapp:2008tf}.
What concerns the free energy for small distances,
it is temperature independent and matches $T=0$
potential, while $r\approx 0.25~\textrm{fm}$
marks the onset of temperature effects.
The temperature dependence of binding energies
obtained with this approach suggests \cite{Jankowski:2009kr} that
charmonium (bottomonium) dissociation takes
place at $T\approx 200~\textrm{MeV}$ ($T\approx 250~\textrm{MeV}$).

%%%%%%%%%%%%%%%%%%%%%%%%%%%%%%%%%

\section{Conclusions}
It is clearly shown that even a crude inclusion
of residual effects from the gluon sector results in a  
remarkable improvement. One of the obvious extensions
of this calculation is to take into account
chiral symmetry effects on the internal quark lines,
governed in this model by the local NJL interaction
\cite{Hansen:2006ee}. In this way one could in principle
study the effects of the chiral transition on $J/\psi$
dissociation.

%%%%%%%%%%%%%%%%%%%%%%%%%%%%%%%%%

\begin{acknowledgments}
We acknowledge inspiring discussions with our colleagues, in particular
K. Redlich, O. Kaczmarek and R. Rapp.
J.J. is grateful for support of his participation at the CPOD 2010 
conference from HIC-for-FAIR and the Bogoliubov-Infeld Programme.
This work was supported in part by the Polish Ministry for Science and 
Higher Education and by the Russian Fund for Basic Research under grant 
No. 11-02-01538-a.
\end{acknowledgments}

\newpage

\newpage 


\begin{thebibliography}{99}


\bibitem{Matsui:1986dk}
\refitem{article}
  T.~Matsui and H.~Satz,
  Phys.\ Lett.\  B {\bf 178} 416 (1986).

\bibitem{Rapp:2008tf}
\refitem{article}
  R.~Rapp, D.~Blaschke and P.~Crochet,
  Prog.\ Part.\ Nucl.\ Phys.\  {\bf 65} (2010) 209
  [arXiv:0807.2470 [hep-ph]].

\bibitem{Riek:2010fk}
\refitem{article}
  F.~Riek and R.~Rapp,
  Phys.\ Rev.\  C {\bf 82} (2010) 035201
  [arXiv:1005.0769 [hep-ph]].

\bibitem{Ebeling:1986}
\refitem{book}
  W.~Ebeling, W.-D.~Kraeft, D.~Kremp, G.~R\"opke,
  {\it Quantum Statistics of Charged Many-Particle Systems},
  Plenum, New York (1986).
  
\bibitem{Megias:2007pq}
\refitem{article}
  E.~Megias, E.~Ruiz Arriola and L.~L.~Salcedo,
  Phys.\ Rev.\  D {\bf 75} (2007) 105019
  [arXiv:hep-ph/0702055].

\bibitem{Jankowski:2009kr}
\refitem{article}
  J.~Jankowski, D.~Blaschke and H.~Grigorian
  Acta Phys.\ Polon.\ Supp.\  {\bf 3} (2010) 747
  [arXiv:0911.1534 [hep-ph]].
  
\bibitem{Kaczmarek:2005gi}
  O.~Kaczmarek and F.~Zantow,
  arXiv:hep-lat/0506019.
  
\bibitem{Ratti:2005jh}
\refitem{article}
  C.~Ratti, M.~A.~Thaler and W.~Weise,
  Phys.\ Rev.\  D {\bf 73} (2006) 014019
  [arXiv:hep-ph/0506234].

\bibitem{Hansen:2006ee}
\refitem{article}
  H.~Hansen, W.~M.~Alberico, A.~Beraudo, A.~Molinari, M.~Nardi and C.~Ratti,
  Phys.\ Rev.\  D {\bf 75} (2007) 065004
  [arXiv:hep-ph/0609116].
  
\bibitem{LeBellac} 
\refitem{book}
M.~LeBellac, {\it Thermal Field Theory}, Cambridge University Press (1996).

\bibitem{Beraudo:2007ky}
\refitem{article}
 A.~Beraudo, J.~P.~Blaizot and C.~Ratti,
  Nucl.\ Phys.\  A {\bf 806} 312 (2008).

\bibitem{Blaschke:2007np}
\refitem{article}
  D.~Blaschke {\it et al.}, %M.~Buballa, A.~E.~Radzhabov and M.~K.~Volkov,
  Yad.\ Fiz.\  {\bf 71}, 2012 (2008).

\bibitem{Petreczky:2004pz}
  P.~Petreczky and K.~Petrov,
    Phys.\ Rev.\  D {\bf 70} (2004) 054503
  [arXiv:hep-lat/0405009].
  
\bibitem{Kaczmarek:2007pb}
  O.~Kaczmarek,
  PoS C {\bf POD07} (2007) 043
  [arXiv:0710.0498 [hep-lat]].

\end{thebibliography}
\end{document}